\documentclass{article}
\usepackage[breaklinks,colorlinks,bookmarks]{hyperref}
\usepackage{fullpage}
\usepackage{latexsym}
\usepackage{amsmath}
\usepackage{amssymb}
\usepackage{amsfonts}

\def\01{\{0,1\}}
\newcommand{\ket}[1]{|#1\rangle}
\newcommand{\bra}[1]{\langle#1|}
\newcommand{\braket}[2]{\langle#1|#2\rangle}
\newcommand{\scalar}[2]{\langle#1,#2\rangle}
\newcommand{\norm}[1]{\|#1\|}

\newcommand{\lmin}{\lambda_{\min}}
\newcommand{\eps}{\varepsilon}
\newcommand{\mhalf}{{-\frac 1 2}}
\newcommand{\good}{\mathrm{good}}
\newcommand{\bad}{\mathrm{bad}}

\newcommand{\ADV}{\mathrm{ADV}}
\newcommand{\MADV}{\mathrm{MADV}}
\newcommand{\dmat}[2]{\begin{pmatrix} #1 & 0 \\ 0 & #2 \end{pmatrix}}

\newcommand{\putabove}[2]{\mathop{#1}\limits^{#2}}
\renewcommand{\i}{\mathbf i}
\newcommand{\Tr}{\mathop{\mathrm{Tr}}}
\newcommand{\nontriv}{2}
\newcommand{\triv}{{\mathrm{triv}}}
\newcommand{\Z}{\mathsf{Z}}
\newcommand{\U}{\mathsf{U}}
\newcommand{\M}{\mathsf{M}}
\newcommand{\F}{\mathsf{F}}
\newcommand{\I}{\mathsf{I}}
\renewcommand{\O}{\mathsf{O}}
\renewcommand{\H}{\mathcal{H}}

\newtheorem{definition}{Definition}
\newtheorem{theorem}{Theorem}
\newtheorem{lemma}[theorem]{Lemma}
\newtheorem{corollary}[theorem]{Corollary}

\newenvironment{proof}[1][Proof]{
    \par
    \noindent \textbf{#1}
}{
    \unskip
    \nobreak\hfill\penalty50\hskip3pt\hbox{}\nobreak\hfill
    \hbox{$\Box$}\par\bigskip
}

\newcommand{\thmref}[1]{\hyperref[#1]{{Theorem~\ref*{#1}}}}
\newcommand{\lemref}[1]{\hyperref[#1]{{Lemma~\ref*{#1}}}}
\newcommand{\corref}[1]{\hyperref[#1]{{Corollary~\ref*{#1}}}}
\newcommand{\eqnref}[1]{\hyperref[#1]{{Eqn.~(\ref*{#1})}}}
\newcommand{\defref}[1]{\hyperref[#1]{{Definition~\ref*{#1}}}}
\newcommand{\secref}[1]{\hyperref[#1]{{Section~\ref*{#1}}}}
\newcommand{\appref}[1]{\hyperref[#1]{{Appendix~\ref*{#1}}}}

\allowdisplaybreaks[1]

\begin{document}
\title{The Multiplicative Quantum Adversary}
\author{%
  Robert \v Spalek%
  \thanks{%
  University of California, Berkeley.  Supported by NSF Grant CCF-0524837
  and ARO Grant DAAD 19-03-1-0082.}\\
  {\tt spalek@eecs.berkeley.edu}
}
\date{}
\maketitle

\begin{abstract}
We present a new variant of the quantum adversary method.  All adversary
methods give lower bounds on the quantum query complexity of a function
by bounding the change of a progress function caused by one query.  All
previous variants upper-bound the \emph{difference} of the progress function,
whereas our new variant upper-bounds the \emph{ratio} and that is why we coin
it the multiplicative adversary.  The new method generalizes to all functions the
new quantum lower-bound method by Ambainis~\cite{ambainis:sdp, asw:symmdpt}
based on the analysis of eigenspaces of the density matrix.  We prove a strong
direct product theorem for all functions that have a multiplicative adversary
lower bound.
\end{abstract}

\section{Introduction}

We consider the problem of proving a lower bound on the number of quantum
queries needed to compute a function with bounded error.  One of the most
successful method for proving quantum query lower bounds is the adversary method
\cite{bbbv:hybrid, ambainis:lowerb, hns:ordered-search, bs:q-read-once, bss:semidef,
ambainis:degree-vs-qc, lm:kolmogorov-lb, zhang:ambainis, ss:adversary,
hls:madv}; see the survey \cite{hs:survey-lb} for the history of the method.
It intuitively works as follows: The computation starts in a fixed quantum state
independent of the input.  The quantum algorithm consecutively applies arbitrary unitary
transformations on its workspace and the input oracle operator.  The quantum state
corresponding to two different inputs $x, y$ gradually diverges to two output states
$\ket{\psi_x^T}, \ket{\psi_y^T}$.  Since
the algorithm has bounded error, there exists a measurement on the output state
that gives the right outcome with high probability, hence the scalar product $|\braket
{\psi_x^T} {\psi_y^T}|$ must be low whenever $f(x) \ne f(y)$ \cite{bv:qctheory}.
We define a \emph{progress function} in time $t$ as a weighted average of these scalar
products over many input pairs:
\begin{equation}
W^t = \sum_{\substack{x,y\\ f(x)\ne f(y)}} w_{x,y} \braket {\psi_x^t} {\psi_y^t} .
\label{eqn:wt}
\end{equation}
Since the scalar products are all one at the beginning and below a constant at the end,
the value of the progress function must drop a lot.  On the other hand, one can show
that one query only causes little \emph{additive} change to the progress function,
hence the algorithm must ask many queries.
The quality of the lower bound depends on the adversary matrix $w$---one typically
has to put more weight on input pairs that are hard to distinguish to get a good bound.

The progress function can be equivalently formulated in terms of density
matrices.  If we run the quantum algorithm on a superposition of
inputs instead of a fixed input, then during the computation the algorithm register
becomes entangled with the
input register.  We trace out the algorithm register and look at the reduced density
matrix $\rho_I^t$ of the input register.  We define the progress function as a linear
form
\begin{equation}
W^t = \scalar \Gamma {\rho_I^t}
\label{eqn:wt2}
\end{equation}
for some Hermitian matrix $\Gamma$ with $\Gamma[x,y]=0$ if $f(x)=f(y)$.  This definition
is equivalent with \eqnref{eqn:wt}
(see \secref{sec:adv}), and since it is easier to work with, we will
stick to it in the whole paper.

The (additive) adversary method suffers one severe limitation: the lower bound
is proportional to the success probability of the algorithm and hence is
negligible for exponentially small success probabilities.  Several many-output functions,
such as $t$-fold search,
however, have strong lower bounds even with exponentially small success
\cite{ksw:dpt-siam} (proved first using the polynomial method \cite{bbcmw:polynomialsj}).
These bounds are useful for proving quantum time-space tradeoffs.

Ambainis reproved \cite{ambainis:sdp} and extended \cite{asw:symmdpt}
these polynomial lower bounds using a new quantum
lower-bound method based on the analysis of subspaces of the reduced density
matrix of the input register.  His method seemed tailored to the problem of quantum
search, and is quite complicated.

In this paper, we reformulate Ambainis's new method in the adversary framework,
generalize it to all functions, and provide some additional intuition.  We use syntactically
the same progress function \eqnref{eqn:wt2}, but in a different way: (1) we require
different conditions on the adversary matrix $\Gamma$, and (2) we show that one query
can only \emph{multiply} the value of the progress function by a small constant---thereout
the name \emph{multiplicative adversary}.  Surprisingly, the final formula for the
multiplicative adversary lower bound is quite similar to the additive adversary.
Unfortunately, the similarity is only illusive, and our new formula is significantly harder to
bound than the old one and we thus don't simplify Ambainis's calculations much.

We, however, split Ambainis's original proof into several independent logical blocks and
dovetail them together: upper-bounding the success probability based on the structure
of the subspaces, upper-bounding the change caused by one query, simultaneous
block-diagonalization of $\Gamma$ and the query operator, and simplification of the final
formula into a form similar to the additive adversary.  This simplification allows us to
generalize the method to all functions.  Furthermore, we separate the
quantum part and the combinatorial part of the proof---the quantum part is hidden
inside the proof of the general lower-bound theorem (\secref{sec:multadv}),
and the user of the method who wants to get a lower bound for some function only has to
evaluate its combinatorial properties (see \secref{sec:examples} for new proofs of all
known bounds in our new framework).

Finally, we show that the multiplicative adversary bound inherently
satisfies the \emph{strong direct product theorem} (DPT).  Roughly speaking it says that
to compute $k$ independent instances of a function we need $\Omega(k)$ times more
queries even if we are willing to decrease the (worst-case) success probability
exponentially.  It is not clear whether this theorem holds for all functions or not.
Ambainis proved a (more complicated) DPT for all symmetric functions directly,
whereas we show that it is sufficient to prove just a (simpler) multiplicative adversary lower
bound, and the DPT then automatically follows (\secref{sec:dpt}).

The biggest open problem, not addressed here, is to find a new stronger lower
bound using the
multiplicative adversary.  A promising function is the element distinctness problem
(tight $\Omega(n^{2/3})$ bound due to the polynomial method \cite{as:collision},
but only $\Omega(\sqrt n)$ adversary lower bound).
It also doesn't seem completely unlikely that one could find a reduction from the additive
adversary to the multiplicative adversary---if a lower bound for some function
can be proved using one method, then it can be proved also using the other one.  If
this is true, then many extensive computations could be avoided.  The multiplicative
adversary method may also have potential to prove a quantum
time-space tradeoff for some Boolean function (again, element distinctness),
because small workspace implies
small Schmidt-rank of the reduced density matrix.
It would be interesting to look at the dual of the multiplicative adversary bound.
The bound is not described by a semidefinite program, however one may be able to use
general Lagrange multipliers.

\section{Adversary framework}
\label{sec:adv}

\subsection{Quantum query complexity}

As with the classical model of decision trees, in the quantum query model
we wish to compute some function $f$ and we access the input through queries.
Let $f: X \rightarrow \Sigma_O$ be a function, with $X \subseteq
\Sigma_I^n$ the set of inputs.  We assume $\Sigma_I = \{ 0, 1, \dots,
\sigma-1 \}$ with $\sigma = |\Sigma_I|$, and call this the input alphabet
and $\Sigma_O$ the output alphabet.
The complexity of $f$ is the number of queries needed to compute $f$ on
a worst-case input $x$.  Unlike the classical case, however, we can
now make queries in superposition.

The memory of a quantum query algorithm is described by three
registers: the input register, $\H_I$, which holds the input $x \in X$,
the query register, $\H_Q$, which holds two
integers $1 \le i \le n$ and $0 \le p < \sigma$, and the working
memory, $\H_W$, which holds an arbitrary value. The query register
and working memory together form the accessible memory, denoted
$\H_A$.

The accessible memory of a quantum query algorithm is initialized to
a fixed state.  For convenience, on input $x$ we assume the state
of the algorithm is $\ket x_I \ket{1,0}_Q \ket 0_W$ where all
qubits in the working memory are initialized to 0.
The state of the algorithm then evolves through queries, which depend on the
input register, and accessible memory operators which do not.
We now describe these operations.

We will model a query by a unitary operator where the oracle answer
is given in the phase.  This operator $\O$ is defined by its action
on the basis state $\ket{x} \ket{i,p}$ as
\[
\O: \ket{x}\ket{i,p} \to e^{\frac {2 \pi \mathbf i} \sigma p x_i}
\ket{x}\ket{i,p},
\]
where $1 \le i \le n$ is the index of the queried input variable and
$0 \le p < \sigma$ is the phase multiplier.  This operation can be
extended to act on the whole space by interpreting it as $\O \times \I_W$,
where $\I_W$ is the identity operation on the workspace $\H_W$.  In the sequel,
we will refer to the action of $\O$ both on $\H_I \otimes \H_Q$ and the full
space $\H_I \otimes \H_Q \otimes \H_W$, and let context dictate which we mean.

For a function with Boolean input $\Sigma_I=\01$, the query operator simply becomes
\[
\O: \ket{x}\ket{i,p} \to (-1)^{p x_i} \ket{x}\ket{i,p},
\]

An alternative, perhaps more common, way to model a quantum query is through
an operator $\O': \ket x \ket{i,p} \to \ket x
\ket{i, (x_i + p) \mod \sigma}$ that encodes the
result in a register.  These two query models are equivalent, as can
be seen by conjugating with the quantum Fourier transform on $\ket p$.
For our results, it is more convenient to work with the phase oracle.

An accessible memory operator is an arbitrary unitary operation $\U$ on the
accessible memory $\H_A$.  This operation is extended to act on the whole space
by interpreting it as $\I_I \otimes \U$, where $\I_I$ is the
identity operation on the input space $\H_I$.
Thus the state of the algorithm on input $x$ after $t$ queries can be written
$$
\ket{\phi_x^t}= \U_t \O \U_{t-1} \cdots \U_1 \O \U_0 \ket x \ket{1,0} \ket 0.
$$
As the input register is left unchanged by the algorithm, we can decompose
$\ket{\phi_x^t}$ as $\ket{\phi_x^t}=\ket{x}\ket{\psi_x^t}$, where
$\ket{\psi_x^t}$ is the state of the accessible memory after $t$ queries.

The output of a $T$-query algorithm on input $x$ is chosen
according to a probability distribution which depends on the final
state of the accessible memory $\ket{\psi_x^T}$.  Namely, the
probability that the algorithm outputs some $b \in \Sigma_O$ on
input $x$ is $\|\Pi_b \ket{\psi_x^T}\|^2$, for a fixed set of
projectors $\{\Pi_b\}$ which are orthogonal and complete, that is,
sum to the identity.  The $\epsilon$-error quantum query complexity
of a function $f$, denoted $Q_{\epsilon}(f)$, is the minimum number
of queries made by an algorithm which outputs $f(x)$ with
probability at least $1-\epsilon$ for every $x$.

\subsection{Progress function}

Imagine that we run some quantum
algorithm on a superposition of inputs $\ket \delta = \sum_{x \in X}
\delta_x \ket x \ket{1,0} \ket 0$.  The quantum state after $t$ queries is
\[
\ket{\Psi^t} = \U_t \O \U_{t-1} \cdots \U_1 \O \U_0 \ket \delta
= \sum_x \delta_x {\ket x} \ket{\psi_x^t} .
\]
The reduced density matrix of the input register is
\[
\rho_I^t
= \Tr_I \ket{\Psi^t} \bra{\Psi^t}
= \sum_{x,y} \delta_x \delta_y^* \braket {\psi_y^t} {\psi_x^t} \cdot \ket x \bra y .
\]

We define a progress function in terms of the reduced density matrix
of the input register and a special Hermitian matrix $\Gamma$, coined the
\emph{adversary matrix}.
We then present two kinds of adversary method, each using the progress
function in a different way to get a quantum query lower bound.

\begin{definition}
Let $\Gamma$ be an $|X| \times |X|$ Hermitian matrix.
Let $\scalar A B = \Tr(A^* B)$.
Define the progress function
\[
W^t = \scalar \Gamma {\rho_I^t} .
\]
Note that $W^t$ is a real number, because both $\Gamma$ and
$\rho_I^t$ are Hermitian.
\end{definition}

In the paper, we will use the following matrices.

\begin{definition}
Define the following set of $|X| \times |X|$ matrices indexed by $i \in [n]$,
$p \in \Sigma_I$, and $z \in \Sigma_O$:
\[
D_i[x,y] = \begin{cases}
    1 & x_i \ne y_i \\
    0 & x_i = y_i \\
\end{cases} ,
\qquad
\O_{i,p}[x,x] = e^{\frac {2 \pi \i} \sigma p x_i} ,
\qquad \mbox{and }
\F_z[x,x] =  \begin{cases}
    1 & f(x) = z \\
    0 & f(x) \ne z \\
\end{cases} .
\]
$D_i$'s are real (zero-one) symmetric matrices. $\O_{i,p}$'s are
diagonal unitary matrices decomposing the query operator $\O =
\bigoplus_{i=1}^n \bigoplus_{p \in \Sigma_I} \O_{i,p}$. $\{ \F_z
\}_{z \in \Sigma_O}$ is a complete set of diagonal orthogonal
projectors, that is $\sum_z \F_z = \I$, $\F_{z_1} \F_{z_2} = 0$ for
$z_1 \ne z_2$, and $\F_z^2 = \F_z$.
\end{definition}

\subsection{Additive adversary}

In this version of the adversary method, one upper-bounds the difference
of the value of the progress function caused by one query.  This method is the
original adversary method, developed in a series of papers \cite{bbbv:hybrid,
ambainis:lowerb, hns:ordered-search, bs:q-read-once, bss:semidef,
ambainis:degree-vs-qc, lm:kolmogorov-lb, zhang:ambainis, ss:adversary,
hls:madv}.

\begin{theorem}[\cite{hls:madv}] \label{thm:additive}
Let $\Gamma$ be a nonzero $|X| \times |X|$ Hermitian matrix such that
$\Gamma[x,y] = 0$ for $f(x) = f(y)$.
Consider a quantum algorithm with error probability at most $\eps$
and query complexity $T$.  We run it on the superposition of inputs $\ket \delta$,
where $\delta$ is a normalized principal eigenvector of $\Gamma$, i.e. corresponding
to the spectral norm $\norm \Gamma$.  Then
\begin{enumerate}
\item
$W^0 = \norm \Gamma$
\item
$W^t - W^{t+1} \le 2 \max_i \norm{\Gamma \circ D_i}$
for every time step $t=0, 1, \dots, T-1$
\item
$W^T \le 2 (\sqrt{\eps (1-\eps)} + \eps) \cdot \norm \Gamma$
\end{enumerate}
\end{theorem}

The third item can be strengthened to $2 \sqrt{\eps (1-\eps)} \norm \Gamma$
if $f$ has Boolean output \cite{hls:madv}.

\begin{corollary}[\cite{bss:semidef, hls:madv}]
\label{cor:addadv}
For every sufficiently small $\eps$,
\[
Q_\eps(f) \ge \ADV_\eps(f)
    \putabove = {\rm def.}
    \left(\frac 1 2 - \sqrt{\eps (1-\eps)} - \eps \right) \max_\Gamma
    \frac {\norm \Gamma} {\max_i \norm{\Gamma \circ D_i}} .
\]
\end{corollary}

If all coefficients of the adversary matrix $\Gamma$ are nonnegative, then
$\Gamma$ corresponds to a hard distribution over input pairs
and its principal eigenvector $\delta$ to a hard distribution over inputs.
The initial value of the progress function $W$ is large, because all scalar products
are one in the beginning, and it must decrease a lot.  This is because the weight is
only put on input pairs evaluating to different outputs, whose scalar product
must be low at the end, otherwise one would not be able to distinguish them.
This intuition does not tell the whole truth if some coefficients are negative, but
the adversary bound still holds.

\section{Multiplicative adversary}
\label{sec:multadv}

In this version of the adversary method, one upper-bounds the ratio
of the value of the progress function before and after a query.  This method
is a simplification and generalization of the new adversary method developed
by Ambainis \cite{ambainis:sdp, asw:symmdpt}.  Here, $\Gamma$ has a
different semantics.  We require the eigenspaces of $\Gamma$ corresponding to small
eigenvalues to be spanned by vectors (superpositions of inputs) that do not
determine the function value with high probability.  The algorithm is then run
on a superposition $\ket\delta$ corresponding to the smallest eigenvalue (which
is typically a uniform superposition of all inputs), and the
progress function $W$ is slowly increasing (instead of decreasing) in time.  To achieve
good success probability, most of the quantum amplitude must move to the higher
subspaces.

\begin{theorem} \label{thm:multadv}
Let $\Gamma$ be positive definite with smallest eigenvalue 1; then $W^t \ge 1$.
Fix a number $1 < \lambda \le \norm\Gamma$.  Let $\Pi_\bad$ be the
projector onto the eigenspaces of $\Gamma$ corresponding to
eigenvalues smaller than $\lambda$.  Assume that
$\norm{\F_z \Pi_\bad}^2 \le \eta$ for every output letter $z \in \Sigma_O$.
Consider a quantum algorithm with success probability
at least $\eta + 4 \zeta$ and query complexity $T$.  We run it on the
superposition of inputs $\ket \delta$, where $\delta$ is a normalized
eigenvector of $\Gamma$ with $\Gamma \delta = \delta$.  Then
\begin{enumerate}
\item
$W^0 = 1$
\item
$\frac {W^{t+1}} {W^t} \le \max_{i,p} \norm{\Gamma_{i,p} / \Gamma}$,
where $\Gamma_{i,p} = \O_{i,p}^* \Gamma \O_{i,p}$
and $\Gamma_{i,p} / \Gamma$ denotes $\Gamma_{i,p} \Gamma^{-1}$
\item
$W^T \ge \zeta^2 \lambda$
\end{enumerate}
\end{theorem}

\begin{corollary}
\label{cor:multadv}
\[
Q_{1 - \eta - 4\zeta} (f) \ge \MADV_{\eta,4\zeta}(f)
\putabove = {\rm def.}
    \max_{\Gamma,\lambda} \frac {\log (\zeta^2 \lambda)}
    {\log(\max_{i,p} \norm{\Gamma_{i,p} / \Gamma})} .
\]
\end{corollary}

\begin{proof}[Proof of \thmref{thm:multadv}(1)]
Trivial, because $\braket {\psi_x^0} {\psi_y^0} = 1$ and thus $W^0 = \scalar
\Gamma {\rho^0_I} = \delta^* \Gamma \delta = 1$.
\end{proof}

\begin{proof}[Proof of \thmref{thm:multadv}(2)]
After the $(t+1)$-st query, the quantum state is $\ket{\Psi^{t+1}} =
\U_{t+1} \O \ket{\Psi^t}$ and thus
\[
\rho^{t+1}_I = \Tr_I (\U_{t+1} \O \ket{\Psi^t} \bra{\Psi^t} \O^*
\U_{t+1}^*) = \Tr_I (\O \ket{\Psi^t} \bra{\Psi^t} \O^*) ,
\]
because the unitary operator $\U_{t+1}$ acts as identity on the
input register. The oracle operator $\O$ only acts on the input
register and the query register, hence we can trace out the working
memory. Denote $\rho = \Tr_{I,Q} \ket{\Psi^t} \bra{\Psi^t}$ and
$\rho' = \O \rho O^*$. Then $\rho^t_I = \Tr_I(\rho)$ and
$\rho^{t+1}_I = \Tr_I(\rho')$.  We re-express the progress function
in terms of $\rho, \rho'$. Define a block-diagonal matrix on
$\H_I \otimes \H_Q$:
\begin{align*}
G &= \Gamma \otimes \I_n \otimes \I_\sigma = \bigoplus
\nolimits_{i=1}^n \bigoplus \nolimits_{p \in \Sigma_I} \Gamma . \\
\intertext{Then}
W^t &= \scalar \Gamma {\rho^t_I} = \scalar G \rho \\
W^{t+1} &= \scalar \Gamma {\rho^{t+1}_I}
= \scalar G {\rho'}
= \scalar G {\O \rho \O^*} \\
&= \scalar {\O^* G \O} \rho
= \scalar {G'} \rho ,
\end{align*}
where $G' = \O^* G \O = \bigoplus_{i,p} \Gamma_{i,p}$ is a block-diagonal
matrix with $\Gamma_{i,p}$'s on the main diagonal.

We upper-bound the change of the progress function as follows.  We show that
\begin{equation}
\scalar {G'} \rho \le \max_{i,p} \norm{\Gamma_{i,p} / \Gamma}
    \cdot \scalar G \rho \label{eq:1q} .
\end{equation}
Since the scalar products $\scalar G \rho$ and $\scalar {G'} \rho$ are linear in $\rho$ and mixed
states are convex combinations of pure states, it suffices to show this
inequality for pure states $\rho = \ket \rho \bra \rho$.
Since both $\Gamma$ and all $\Gamma_{i,p}$'s are positive
definite, both $G$ and $G'$ are also positive definite.
Let $\ket \tau = \sqrt G \ket \rho$, that is $\ket \rho = G^\mhalf \ket \tau$.
\begin{align*}
\frac {\scalar {G'} \rho} {\scalar G \rho}
&= \frac {\bra \rho G' \ket \rho} {\bra \rho G \ket \rho}
= \frac {\bra \tau G^\mhalf G' G^\mhalf \ket \tau} {\braket \tau \tau}
 = \left( \frac {\norm{\sqrt{G' / G} \ket \tau}} {\norm \tau} \right)^2 \\
&\le \norm{\sqrt{G' / G}}^2 = \norm{G' / G}
= \max_{i,p} \norm{\Gamma_{i,p} / \Gamma}
\end{align*}
We conclude that \eqnref{eq:1q} holds for pure states and consequently also for all
density matrices.
\end{proof}

\goodbreak

Since the spectral norm of $\Gamma_{i,p} / \Gamma$ is hard to
upper-bound, we simplify it to a form similar to the additive adversary.

\begin{lemma} \label{lem:diagonal}
Fix the index of the queried bit $i$. Assume that $\Gamma$ and
$\O_{i,1}$ are simultaneously block-diagonal in some ``basis'', that
is there exists a complete set of orthogonal projectors $\Pi = \{
\Pi_\ell \}_\ell$ such that $\Gamma = \sum_\ell \Gamma^{(\ell)}$ and
$\O_{i,1} = \sum_\ell \O_{i,1}^{(\ell)}$, where $\Gamma^{(\ell)}$ denotes
$\Pi_\ell \Gamma \Pi_\ell$.  Note that since $\O_{i,p} = (\O_{i,1})^p$,
all $\O_{i,p}$ are block-diagonal in this basis, too.  Then
\[
\norm{\Gamma_{i,p} / \Gamma}
    = \max_\ell \norm {\Gamma_{i,p}^{(\ell)} / \Gamma^{(\ell)}}
    \le 1 + 2 \max_\ell \frac
    {\norm{\Gamma^{(\ell)} \circ D_i}} {\lmin(\Gamma^{(\ell)})} ,
\]
where $\lmin(M)$ denotes the smallest eigenvalue of $M$.
\end{lemma}

\begin{corollary}
\label{cor:multadv2}
Let $\Gamma \succeq \I$ and $1 < \lambda \le \norm \Gamma$.  Assume that
$\norm{\F_z \Pi_\bad}^2 \le \eta$ for every $z \in \Sigma_O$, where $\Pi_\bad$
is the projector onto the eigenspaces of $\Gamma$ smaller than $\lambda$.
For a fixed $i$, block-diagonalize simultaneously $\Gamma$ and $\O_{i,1}$,
and let $\Gamma^{(\ell)}$ denote the $\ell$-th block.
\[
\MADV_{\eta, 4\zeta}(f) \ge \max_{\Gamma,\lambda} \log(\zeta^2 \lambda)
    \cdot \min_{i,\ell} \frac {\lmin(\Gamma^{(\ell)})} {2 \norm{\Gamma^{(\ell)} \circ D_i}} .
\]
\end{corollary}

Note that for most functions, one typically does not want to apply this
statement using the trivial block-diagonalization $\Pi = \{ \I \}$ with just
one projector onto the whole space.  In this case, $\norm{\Gamma
\circ D_i}$ is way too large compared to $\lmin(\Gamma)$
and the final bound is too weak.  For some functions, such as
unordered search, however, even this approach gives a reasonable bound;
see \lemref{lem:search}.

\begin{proof}[Proof of \lemref{lem:diagonal}]
Denote $Y_{i,p} = \O_{i,p}^* E \O_{i,p}$, where $E$ is the all-ones
matrix; then $\Gamma_{i,p} = \O_{i,p}^* \Gamma \O_{i,p} = \Gamma \circ
Y_{i,p}$.  Note that $Y_{i,0} - \frac 1 \sigma \sum_{p=0}^{\sigma-1}
Y_{i,p} = D_i$ (by summing a geometrical sequence depending on $x_i-y_i$).
Since $\Gamma$ and all $\O_{i,p}$ are block-diagonal
in $\Pi$, it follows that $\Gamma_{i,p}$, $\Gamma_{i,p} / \Gamma$,
and $\Gamma \circ D_i$ are block-diagonal in $\Pi$, too.  Therefore
it suffices to compute an upper bound on $\norm{\Gamma_{i,p} /
\Gamma}$ in each subspace of $\Pi$ separately and take the maximum
as the total upper bound.  This proves the first part of the lemma that
$\norm{\Gamma_{i,p} / \Gamma} = \max_\ell \norm {\Gamma_{i,p}^{(\ell)} /
\Gamma^{(\ell)}}$.

Henceforth, fix the index $\ell$ of one such
subspace and let $\Gamma := \Gamma^{(\ell)}$ denote the matrix projected onto
$\Pi_\ell$.
\begin{align*}
\norm {\Gamma_{i,p} \Gamma^{-1}}
&= \max_{v \in \Pi_\ell} \frac {\norm{\Gamma_{i,p} \Gamma^{-1} v}} {\norm v}
    && v = \Gamma w \\
&= \max_{w \in \Pi_\ell} \frac {\norm{\Gamma_{i,p} w}} {\norm {\Gamma w}} \\
\frac {\norm{\Gamma_{i,p} w}} {\norm {\Gamma w}}
&= \frac {\norm{\Gamma w + (\Gamma_{i,p} - \Gamma) w}} {\norm {\Gamma w}} \\
&\le 1 + \frac {\norm{(\Gamma_{i,p} - \Gamma) w}}
    {\norm {\Gamma w}}
    && \norm{\Gamma w}
    \ge \underbrace{\lmin(\Gamma)}_\mu \norm w \\
&\le 1 + \frac 1 \mu \cdot \frac {\norm{(\Gamma_{i,p} - \Gamma) w}} {\norm w}
    && \Gamma_{i,p} - \Gamma = (\Gamma_{i,p} - \Gamma) \circ D_i \\
&= 1 + \frac 1 \mu \cdot \frac {\norm{(\Gamma_{i,p} \circ D_i) w
    - (\Gamma \circ D_i) w}} {\norm w}
    && \Gamma_{i,p} = \Gamma \circ Y_{i,p} \\
&\le 1 + \frac 1 \mu \cdot \frac {\norm{(\Gamma \circ Y_{i,p} \circ D_i) w}
    + \norm{(\Gamma \circ D_i) w}} {\norm w}
    && (\Gamma \circ D_i) \circ Y_{i,p} = \O_{i,p}^* (\Gamma \circ D_i) \O_{i,p} \\
&\le 1 + \frac {\norm{\O_{i,p}^* (\Gamma \circ D_i) \O_{i,p}}
    + \norm{\Gamma \circ D_i}} {\mu}
    && \O_{i,p} \mbox{ is unitary} \\
&= 1 + 2 \frac {\norm{\Gamma \circ D_i}} {\lmin(\Gamma)} ,
\end{align*}
which proves the second part of the lemma, because $\Gamma$ here denotes
$\Gamma^{(\ell)}$.
\end{proof}

\begin{proof}[Proof of \thmref{thm:multadv}(3)]
At the end of the computation, we measure the input register in the
computational basis and the accessible memory according to the
output projectors $\{ \Pi_b \}$.  Denote the outcomes $x \in X$ and
$b \in \Sigma_O$. Since the algorithm has good success probability,
$f(x)=b$ with probability at least $\eta + 4 \zeta$.  Let us prove
an upper bound on this success probability in terms of the progress
function.

Let $\Pi_\good = \I - \Pi_\bad$ denote the projector onto the
orthogonal complement of the bad subspace, coined the good subspace.
We upper-bound the success probability in the bad subspace by $\eta$
and in the good subspace by 1.  Consider the final state of the
computation $\ket {\Psi^T}$; recall that $\rho_I^T = \Tr_I \ket{\Psi^T}
\bra{\Psi^T}$. Let $\ket{\Psi_\bad} = \frac {\Pi_\bad \ket {\Psi^T}}
{\norm{\Pi_\bad \ket {\Psi^T}}}$, $\ket{\Psi_\good} = \frac
{\Pi_\good \ket {\Psi^T}} {\norm{\Pi_\good \ket {\Psi^T}}}$, and
$\beta = \scalar {\Pi_\good} {\rho_I^T} = \norm{\Pi_\good \ket
{\Psi^T}}^2$.  (When using a projector on a larger Hilbert space
than defined, we first extend it by a tensor product with identity.
For example, $\ket{\Psi_\bad} = \frac {(\Pi_\bad \otimes \I_A) \ket
{\Psi^T}} {\norm{(\Pi_\bad \otimes \I_A) \ket {\Psi^T}}}$, where $A$
is the accessible memory.) Decompose
\[
\ket {\Psi^T} = \sqrt{1-\beta} \ket{\Psi_\bad} + \sqrt \beta
\ket{\Psi_\good} .
\]
Assume for a moment that the final state was $\ket{\Psi_\bad}$
instead of $\ket{\Psi^T}$.  We measure the accessible memory first
and fix the output of the computation $b \in \Sigma_O$, then we
trace out the accessible memory completely and end up with a mixed
state $\rho$ over the input register (not necessarily equal to
$\rho_I^T$, because we remember $b$). We then measure the input
register according to the projectors $\{ \F_z \}$ (set of inputs $x$
such that $f(x)=z$) and test whether $z = b$. Now, for every $z \in
\Sigma_O$, including the right result $z = b$,
\begin{align*}
\Pr[\mbox{obtaining }z] &= \scalar {\F_z} \rho
    && \mbox{$\rho$ is only supported on $\Pi_\bad$} \\
&= \scalar {\F_z} {\Pi_\bad \rho \Pi_\bad} \\
&= \scalar {\Pi_\bad \F_z \Pi_\bad} \rho
    && \scalar A B \le \norm A \cdot \norm B_{tr} \\
&\le \norm{\Pi_\bad \F_z \Pi_\bad} \cdot \norm \rho_{tr}
    && \norm{\rho}_{tr}=1 \\
&= \norm{\Pi_\bad \F_z \Pi_\bad}
    && \F_z = \F_z^2 \\
&= \norm{\Pi_\bad \F_z \cdot \F_z \Pi_\bad}
    && \norm {A \cdot B} \le \norm A \cdot \norm B \\
&\le \norm{\Pi_\bad \F_z} \cdot \norm{\F_z \Pi_\bad}
    && \F_z, \Pi_\bad \mbox{ are Hermitian} \\
&= \norm{\F_z \Pi_\bad}^2 \\
&\le \eta .
\end{align*}
Therefore the success probability of the algorithm would be at most
$\eta$, had the input register been in the state $\ket{\Psi_\bad}$.
The real output state is $\ket {\Psi^T}$.  Since the trace distance
of these two states is
\[
\norm{\ket {\Psi^T} - \ket {\Psi_\bad}}
  \le (1 - \sqrt{1-\beta}) + \sqrt \beta \le 2 \sqrt \beta ,
\]
by \cite{bv:qctheory}, the success probability on $\ket {\Psi^T}$
could be at most $\eta + 4 \sqrt \beta$.  On the other hand, we
assumed that the algorithm has success probability at least $\eta +
4 \zeta$, hence $\beta \ge \zeta^2$. The progress function at the
end takes value
\[
W^T = \scalar \Gamma {\rho_I^T} \ge \scalar {\lambda \cdot
\Pi_\good} {\rho_I^T} = \beta \lambda \ge \zeta^2 \lambda ,
\]
which is what we had to prove.
\end{proof}

\section{Applications}
\label{sec:examples}

In this section, we reprove all known bounds obtained by the subspace-analysis
technique of Ambainis.  We only consider functions with Boolean input.  The input oracle rotates
the phase by a factor of $(-1)^{p x_i}$ and the only nontrivial case is $p=1$.
We thus omit $p$ and write just $\O_i, \Gamma_i$ instead of $\O_{i,1},
\Gamma_{i,1}$.

\subsection{Search}
\label{sec:search}

\newcommand{\search}{\mathrm{Search}_n}
Let $X = \{ x \in \01^n: |x| = 1\}$ and $\search(x) = i$ such that $x_i=1$.  In
other words, there is exactly one 1 in an $n$-bit string and we have
to find it.  One can quickly estimate the multiplicative adversary
bound as follows.

\begin{lemma}
\label{lem:search}
$\MADV_{n^{-1},4\zeta}(\search) = \Omega(\zeta^2 \sqrt n)$.
\end{lemma}

\begin{proof}
Let $q > 1$ be a constant whose value we fix later. Define the
following unit vectors: $v = \frac 1 {\sqrt n} (1, \dots, 1)$ and
$v_i = \frac 1 {\sqrt{n (n-1)}} (1, \dots, 1, 1-n, 1, \dots, 1)$
with $1-n$ on the $i$-th position.  Note that $v \perp v_i$, but
$v_i \not\perp v_j$ for $i \ne j$.  Define the following adversary
matrix:
\begin{equation}
\Gamma = (1-q) \ket v \bra v + q \I ,
\label{eqn:gammasearch}
\end{equation}
where $\I$ is the identity matrix. $\Gamma$ has two eigenspaces:
$\Gamma v = v$, and $\Gamma v_i = q v_i$.
The success probability in the subspace of $v$ is $\eta = 1/n$.
Let $\lambda = \norm \Gamma = q$.

We apply \corref{cor:multadv2} with trivial block-diagonalization
$\Pi = \{ \I \}$.  Then $\lmin(\Gamma)=1$.  $\Gamma \circ D_i$ consists
of an $1 \times (n-1)$ block $\frac {1-q} n (1, \dots, 1)$ and its adjoint, hence
$\norm{\Gamma \circ D_i} = \sqrt{n-1} \cdot (q-1) / n < (q-1) / \sqrt n$.  Hence
\begin{equation}
\MADV_{n^{-1},4\zeta}(\search) \ge \frac {\log(\zeta^2 q)} {2 (q-1)} \sqrt n .
\label{eqn:search1}
\end{equation}
We set $q = 2 / \zeta^2$ to make the logarithm positive.
\end{proof}

It turns out that the rough analysis in the previous lemma loses a quadratic factor
in the success probability.  Let us compute exactly the eigenvalues of
$\Gamma_i \Gamma^{-1}$.  Thanks to the symmetry, it
is sufficient to only consider one case $i=1$.

\begin{theorem}
$\MADV_{n^{-1},4\zeta}(\search) = \Omega(\zeta \sqrt n)$.
\end{theorem}

\begin{proof}
We use the same adversary matrix \eqnref{eqn:gammasearch}.  We could
compute $\Gamma_1 \Gamma^{-1}$ explicitly, but we instead choose to
demonstrate the block-diagonalization process.  Define a complete
set of orthogonal projectors $\Pi = \{ \Pi_\nontriv, \Pi_\triv \}$
with a 2-dimensional subspace $\Pi_\nontriv = \ket v \bra v + \ket
{v_1} \bra {v_1}$ and its orthogonal complement $\Pi_\triv = \I -
\Pi_\nontriv$.  Define $\ket {w_2} = \Pi_\triv \ket{v_2} = \ket{v_2}
- \braket {v_1} {v_2} \ket {v_1} = \sqrt{ \frac n {(n-1)^3} } (0,
2-n, 1, \dots, 1)$ for which $v_1 \perp w_2$, and define similarly
$w_3, \dots, w_n$.  Then $\Pi_\triv$ is spanned by $w_2, \dots,
w_n$.  $\O_1 w_i = w_i$ implies $\Pi_\triv \O_1 = \Pi_\triv$. Since
$\O_1$ is unitary, $\O_1$ is block-diagonal in $\Pi$.  Now,
\[
\Pi_\nontriv \Gamma = \ket v \bra v + q \ket {v_1} \bra {v_1}
\]
and $\Pi_\triv \Gamma = q (\I - \ket v \bra v - \ket {v_1} \bra
{v_1}) = q \Pi_\triv$, and hence $\Gamma$ is also block-diagonal in
$\Pi$.  We now analyze the diagonal blocks of $\Gamma_1 \Gamma^{-1}$
in this ``basis''.  We already know that $\Pi_\triv \O_1 = \Pi_\triv$
and hence $\Gamma_1 \Gamma^{-1} = \I$ on the trivial subspace.  It
remains to examine the non-trivial subspace $\Pi_\nontriv$.

In the orthonormal basis $\{ \ket v, \ket {v_1} \}$,
\[
\Gamma = \begin{pmatrix} 1 & 0 \\ 0 & q \end{pmatrix} ,
\qquad \O_1 = \frac 1 n \begin{pmatrix}
n-2 & 2 \sqrt{n-1} \\
2 \sqrt{n-1} & 2 - n
\end{pmatrix} ,
\]
and the eigenvalues of $\Gamma_1 \Gamma^{-1} = \O_1^* \Gamma \O_1
\Gamma^{-1}$ are $1 \pm \frac {2 (q-1)} {\sqrt{q n}} + O(\frac 1 n)$.
Hence $\norm {\Gamma_1 / \Gamma} \approx 1 + \frac {2 (q-1)} {\sqrt{q n}}$,
and by \corref{cor:multadv}, the multiplicative adversary bound is
\[
\MADV_{n^{-1}, 4\zeta}(\search)
    \ge \frac {\log (\zeta^2 \lambda)} {\log (1 + 2 (q-1) / \sqrt{q n})}
    \ge \frac {\log(\zeta^2 q) \sqrt q} {2 (q-1)} \sqrt n ,
\]
where we have used that $\log(1+x) \le x$.  This is by a factor of $\sqrt q$ larger than
the bound given by \eqnref{eqn:search1}.  Again, we set $q = 2/\zeta^2$ and finish
the proof.
\end{proof}

\subsection{$t$-fold search}

\newcommand{\tsearch}{\mathrm{Search}_{t,n}}
This is a generalization of the search problem, where we have to find $t$ ones.
Let $X = \{ x \in \01^n: |x|=t \}$ and $\tsearch(x)=J$ such that $J \subseteq [n]$, $|J|=t$,
and $x_J=1$.
The additive adversary implies that the bounded-error quantum query complexity
of $\tsearch$ is $\Omega(\sqrt{t n})$.  The multiplicative adversary gives the same bound
even for an exponentially small success probability!  This whole section is based
on the analysis by Ambainis~\cite{ambainis:sdp} translated to our framework.

\begin{theorem}[\cite{ambainis:sdp}] \label{thm:tsearch}
For every $t \le \frac n {4 e}$, $\MADV_{2^{-t/2}, 2^{-t/8}}(\tsearch) = \Omega(\sqrt{t n})$.
\end{theorem}

\begin{proof}
Fix a $1 < q < O(1)$ and set $\lambda = q^{t/2}$.  Define the following adversary matrix:
\[
\Gamma = \sum_{j=0}^t q^j \Pi_{S_j} ,
\]
where $\Pi_S$ denotes the projector onto the subspace $S$,
$S_j = T_j \cap T_{j-1}^\perp$, and $T_j$ is the space spanned by
\[
\ket{\psi_J} = \frac 1 {\sqrt{{n-j \choose t-j}}}
    \sum_{\substack{x: |x|=t \\ x_J=1}} \ket x
    \qquad \mbox{for $J \subseteq [n]$ with $|J|=j$.}
\]
Denote $\ket{\tilde\psi_J} = \Pi_{T_{j-1}^\perp} \ket{\psi_J}$.
These projected states are neither normalized nor orthogonal for $j
> 0$.  Denote $\ket{\ddot\psi_J} = \frac {\ket{\tilde\psi_J}}
{\norm{\tilde\psi_J}}$.  Note that $S_j$ is spanned by
$\ket{\ddot\psi_J}$.

\paragraph{Block-diagonalization of $\Gamma$ and $\O_{i,p}$}
Thanks to the symmetry, it is sufficient to only consider the case
$i=1$ of querying the first input bit.  As we say above, the only
nontrivial case is $p=1$.  We present a complete set of orthogonal
projectors $\Pi$ in which both $\Gamma$ and $\O_1$ are
block-diagonal.  Let
\begin{align*}
\ket{\psi_J^b} &= \frac 1 {\sqrt{{n-j-1 \choose t-j-b}}}
    \sum_{\substack{x: |x|=t \\ x_1 = b\\ x_J=1}} \ket x
    && \mbox{for $J \subseteq [n]$ such that $1 \not\in J$} \\
\ket{\tilde\psi_J^b} &= \Pi_{T_{j-1,b}^\perp} \ket{\psi_J^b}
    && \mbox{with $T_{j,b}$ spanned by $\ket{\psi_J^b}$ with $|J|=j$} \\
\ket{\ddot\psi_J^b} &= \frac {\ket{\tilde\psi_J^b}} {\norm{\tilde\psi_J^b}}
\end{align*}
Let $S_{j,b} = T_{j,b} \cap T_{j-1,b}^\perp$.  Then the following
holds:
\begin{itemize}
\item
Let $\ket{\ddot\psi_J^{a,b}}$ denote the vector $a
\ket{\ddot\psi_J^0} + b \ket{\ddot\psi_J^1}$. Let
\begin{align}
\alpha'_j &= \sqrt{\frac {n-t} {n-j}} \norm{\tilde\psi_J^0} &
\beta'_j &= \sqrt{\frac {t-j} {n-j}} \norm{\tilde\psi_J^1} \nonumber \\
\alpha_j &= \frac {\alpha_j'} {\sqrt{(\alpha_j')^2 + (\beta_j')^2}}
& \beta_j &= \frac {\beta_j'} {\sqrt{(\alpha_j')^2 + (\beta_j')^2}}
\label{eqn:ab}
\end{align}
We also denote them $\alpha, \beta$ if the index $j$ is clear from
the context.  Note that $\alpha^2 + \beta^2 = 1$.  Then
$\ket{\ddot\psi_J^{\alpha, \beta}} \in S_j$ and
$\ket{\ddot\psi_J^{\beta, -\alpha}} \in S_{j+1}$ \cite[Claim
15]{asw:symmdpt}.  These two new vectors span the same subspace as
$\ket{\ddot\psi_J^0}$ and $\ket{\ddot\psi_J^1}$.
\item \cite[Claim 16]{asw:symmdpt}
$S_{j,0}$ and $S_{j,1}$ have the same dimension and the mapping
\[
M' \ket{0 x_2 \dots x_n} \to \sum_{\ell: x_\ell=1} \ket{1 x_2 \dots
x_{\ell-1} 0 x_{\ell+1} \dots x_n}
\]
is a multiple $M' = c_j \M_j$ of some unitary operation on $S_{j,0}
\to S_{j,1}$ that maps $\M_j: \ket{\ddot\psi_J^0} \to
\ket{\ddot\psi_J^1}$.
\item
Pick any orthonormal basis $\{ \ket{\varphi_{j,\ell}} \}_\ell$ for
each $S_{j,0}$ (the defining basis $\ket{\ddot\psi_J^0}$ is not
orthogonal).  For $j<t$, define projectors $\Pi_{j,\ell} =
\ket{\varphi_{j,\ell}} \bra{\varphi_{j,\ell}} + \M_j
\ket{\varphi_{j,\ell}} \bra{\varphi_{j,\ell}} \M_j^*$.  Note that if
some $\ket{\varphi_{j,\ell}} = \ket{\ddot\psi_J^0}$, then
\begin{equation}
\Pi_{j,\ell}
= \ket{\ddot\psi_J^0} \bra{\ddot\psi_J^0}
+ \ket{\ddot\psi_J^1} \bra{\ddot\psi_J^1}
= \ket{\ddot\psi_J^{\alpha,\beta}} \bra{\ddot\psi_J^{\alpha,\beta}}
+ \ket{\ddot\psi_J^{\beta,-\alpha}} \bra{\ddot\psi_J^{\beta,-\alpha}}
\label{eqn:searchproj}
\end{equation}
due to the basis change mentioned in the first item above.  Thus
$\Pi_j = \{ \Pi_{j,\ell} \}_\ell$ is a complete set of orthogonal
projectors for $S_{j,0} \oplus S_{j,1}$, or, equivalently, for the
subspace of $S_j \cup S_{j+1}$ spanned by $\ket{\tilde\psi_J}$ and
$\ket{\tilde\psi_{J \cup \{1\}}}$ with $|J|=j$ and $1 \not\in J$. It
follows that
\[
\Pi = \underbrace{\{ \Pi_{j,\ell} \}_{j,\ell}}_{\mbox{2-dim
projectors}} \cup \underbrace{\{ \ket{\varphi_{t,\ell}}
\bra{\varphi_{t,\ell}} \}_\ell}_{\mbox{trivial subspace $S_{t,0}$}}
\]
is a complete set of orthogonal projectors for the whole input space
$T_t$.
\end{itemize}

Let us verify that $\Pi$ indeed block-diagonalizes $\Gamma$ and
$\O_1$. To compute the images of the basis states of each projector
$\Pi_{j,\ell}$, we use a double decomposition like in
\eqnref{eqn:searchproj}.  First, since $\ket{\varphi_{j,\ell}} \in
S_{j,0}$ and $\M_j \ket{\varphi_{j,\ell}} \in S_{j,1}$, $\O_1
\ket{\varphi_{j,\ell}} = \ket{\varphi_{j,\ell}}$ and $\O_1 \M_j
\ket{\varphi_{j,\ell}} = -\M_j \ket{\varphi_{j,\ell}}$, hence $\O_1$
is block-diagonal in $\Pi$. Second, if we denote
$\ket{\varphi_{j,\ell}^{a,b}} = a \ket{\varphi_{j,\ell}} + b \M_j
\ket{\varphi_{j,\ell}}$, then $\ket{\varphi_{j,\ell}
^{\alpha,\beta}} \in S_j$ and $\ket{\varphi_{j,\ell}
^{\beta,-\alpha}} \in S_{j+1}$, because both
$\ket{\varphi_{j,\ell}}$ and $\M_j \ket{\varphi_{j,\ell}}$ are just
linear combinations with the same coefficients of states
$\ket{\ddot\psi_J^0}$ and $\ket{\ddot\psi_J^1}$ respectively.  We
conclude that $\Gamma \ket{\varphi_{j,\ell}^{\alpha,\beta}} = q^j
\ket{\varphi_{j,\ell}^{\alpha,\beta}}$ and $\Gamma
\ket{\varphi_{j,\ell}^{\beta,-\alpha}} = q^{j+1}
\ket{\varphi_{j,\ell}^{\beta,-\alpha}}$, hence $\Gamma$ is also
block-diagonal in $\Pi$.

\paragraph{Eigenvalues of $\Gamma_1 / \Gamma$ on $\Pi_{j,\ell}$}
Consider the orthonormal basis $B_1 = \{ \ket{\varphi_{j,\ell}}, \M_j
\ket{\varphi_{j,\ell}} \}$ of $\Pi_{j,\ell}$.  Let $\U =
(\begin{smallmatrix} \alpha & \beta \\ \beta & -\alpha
\end{smallmatrix})$ denote the self-adjoint unitary operator changing
the basis from $B_2 = \{ \ket{\varphi_{j,\ell}^{\alpha,\beta}},
\ket{\varphi_{j,\ell}^{\beta,-\alpha}} \}$ to $B_1$. Then, in the basis
$B_1$,
\begin{align*}
\Gamma_1 \Gamma^{-1} \Pi_{j,\ell}
    &= \dmat 1 {-1} \underbrace {\U \dmat {q^j} {q^{j+1}} \U^*}_\Gamma \dmat 1 {-1}
    \underbrace {\U \dmat {q^{-j}} {q^{-j-1}} \U^*}_{\Gamma^{-1}} \\
&= \left(
    \left( 1 + 2 \alpha^2 \beta^2 \frac {(q-1)^2} q
    \right) \I + 2 \alpha \beta \frac {q-1} q
    \begin{pmatrix} 0 & \alpha^2 + \beta^2 q \\
    \beta^2 + \alpha^2 q & 0 \end{pmatrix}
    \right).
\end{align*}
A straightforward calculation shows that $\Gamma_1 \Gamma^{-1}
\Pi_{j,\ell}$ has eigenvalues
\[
1 + 2 \alpha^2 \beta^2 \frac {(q-1)^2} q
    \pm 2 \alpha \beta \frac {q-1} q
    \sqrt{(\alpha^2 + \beta^2 q) (\beta^2 + \alpha^2 q)} .
\]
We use a trivial upper bound $\alpha \le 1$, and an upper bound $\beta \le \sqrt{2 t / n}$
\cite[Claim 19]{asw:symmdpt}, which easily follows from the (non-trivial) computation of
$\norm{\tilde\psi_J^b} = \sqrt{ \frac {(n-t+b-1)^{\underline j}} {(n-j)^{\underline j}} }$
\cite[Claim 18]{asw:symmdpt}.  By using $1 < q \le n / t$, and $q = O(1)$, we obtain
\[
\norm{\Gamma_1 \Gamma^{-1} \Pi_{j,\ell}}
    \le 1 + \frac {4 t (q-1)^2} {n q}
    + 2 \sqrt{\frac {2 t} n} \frac {q-1} q \sqrt{3 (q + 2 t / n)}
    < 1 + 6 \sqrt{ \frac {2 t} {n q} } (q-1) + O(\tfrac t n) .
\]

\paragraph{Eigenvalues of $\Gamma_1 / \Gamma$ on the trivial subspace}
Let us revisit the trivial subspace $S_{t,0}$ and make sure that the
block-diagonalization is right there, too.  We claim that $S_{t,0}
\subseteq S_t$.  Since $\ket{\psi_J^0} = \ket{\psi_J}$ for $|J|=t$
and $1 \not\in J$, we get $T_{t,0} \subseteq T_t$, and thus it
suffices to prove $S_t = T_{t,0} \cap T_{t-1,0}^\perp \subseteq
T_{t-1}^\perp$.  It holds that $T_{t,0} \subseteq T_{t-1,1}^\perp$
due to a different value of the first input bit. Also, we know that
$T_j \subseteq T_{j,0} \oplus T_{j,1}$, hence $T_{j,0}^\perp \cap
T_{j,1}^\perp \subseteq T_j^\perp$ and the proof is finished.

Let $\ket w \in S_{t,0}$.  $\ket w$ is an eigenvector of $\Gamma$,
because it lies in $S_t$.  Since $\O_1 \ket w = \ket w$, we conclude
that $\Gamma_1 \Gamma^{-1} \ket w = \ket w$ and $S_{t,0}$ is indeed
a trivial subspace.

\paragraph{Upper-bounding $\eta$}
Since $\lambda = q^{t/2}$, we have to upper-bound $\norm{\F_z \ket\varphi}^2 \le \eta$
for all $\ket\varphi \in T_{t/2}$.  Since the dimension of $T_{t/2}$ is $n \choose t/2$ and
the number of possible outcomes is $n \choose t$, using \cite{nayak:decoding}, the
success probability is at most $\eta \le {n \choose t/2} / {n \choose t}$.  Using the
bounds $(\frac n k)^k \le {n \choose k} \le (e \frac n k)^k$ and assuming
$t \le \frac n {4 e}$,
\[
\eta
\le \frac {{n \choose t/2}} {{n \choose t}}
\le \frac {(2 e \frac n t)^{t/2}} {(\frac n t)^t}
= \left( \frac {2 e t} n \right)^{t/2}
\le 2^{-t/2} .
\]
By being more careful, one can prove an exponentially small upper bound on
$\eta$ for all $t \le \frac n 2$.

\paragraph{Multiplicative adversary bound}
Since $\lambda = q^{t/2}$ and $\norm{\Gamma_1 / \Gamma} = \max_j
\norm{\Gamma_1 \Gamma^{-1} \Pi_{j,\ell}}$, by \corref{cor:multadv},
the multiplicative adversary bound is
\[
\MADV_{\eta, 4\zeta}(\tsearch) \ge \frac {\log(\zeta^2 q^{t/2})} {\log(1 + 6 (q-1) \sqrt{2 t / n q})}
    \ge \frac {\frac t 2 \log q + 2 \log \zeta} {6 (q-1) \sqrt{2 t / q}} \sqrt{n}
    = \frac {\log q - 4 \log(\zeta^{-1/t})} {12 (q-1) \sqrt{2 / q}} \sqrt{t n} ,
\]
which is $\ge \frac {\log 2} {24} \sqrt{t n}$ for $q=2$ and $\zeta \ge 2^{-t/8}$.
\end{proof}

\subsection{$t$-threshold function}

\newcommand{\tthreshold}{\mathrm{Threshold}_{t,n}}
The decision version of the $t$-fold search problem is the
$t$-threshold function $X = \{ x \in \01^n: |x| \in \{ t-1, t \} \}$
and $\tthreshold(x) = |x|-t+1$. Here one can always achieve success
probability $1/2$ by random guess, hence we want to upper-bound the
bias from $1/2$.  The analysis in this section is based on Ambainis's
method \cite{asw:symmdpt} translated to our framework.

\begin{theorem}[\cite{asw:symmdpt}]
\label{thm:threshold}
$\MADV_{1/2, 4 \zeta}(\tthreshold) = \Omega(\zeta^2 \sqrt{t n})$.
\end{theorem}

One may think that the true bound is $\Omega(\zeta \sqrt{t
n})$, however we are unable to prove it using this method.  It is quite hard to analyze
the $4 \times 4$ matrix in the following proof exactly, and we rather use the simpler
bound from \corref{cor:multadv2}, which loses exactly this quadratic factor
in \lemref{lem:search}.  We tried to do exact calculations in Mathematica,
but they seem to give the same bound $\Omega(\zeta^2 \sqrt{t n})$ even when using
\corref{cor:multadv}.

\begin{proof}[Proof (sketch)]
We conduct the proof similarly to \thmref{thm:tsearch}, but now we
use eigenspaces spanned by uniform superpositions of \emph{both}
$(t-1)$-weight and $t$-weight strings.  Define the following
adversary matrix with $q = 1 + \frac {4 \log(2/\zeta)} t$ and
$\lambda = q^{t/2}$:
\begin{equation}
\Gamma = \underbrace{\sum_{j=0}^{t/2-1} q^j \Pi_{S_{j,+}}}_\bad
    + q^{t/2} \underbrace{\left( \sum_{j=t/2}^{t-1} \Pi_{S_{j,+}}
    + \sum_{j=0}^t \Pi_{S_{j,-}} \right)}_\good ,
    \label{eqn:gamma-thr}
\end{equation}
where $S_{j,\pm}$ is spanned by $\ket{\ddot\psi_{J,\pm}} = \frac 1
{\sqrt 2} (\ket{\ddot\psi_{J,0}} \pm \ket{\ddot\psi_{J,1}})$,
\[
\ket{\psi_{J,a}} = \frac 1 {\sqrt{{n-j \choose t-1+a-j}}}
    \sum_{\substack{x: |x|=t-1+a \\ x_J=1}} \ket x ,
\]
and the tilde and double-dot states are defined as usual.

\goodbreak

Let us explain the intuition behind this construction.  We have to
put all minus subspaces inside the good subspace of $\Gamma$,
otherwise some $\ket v = \ket{\ddot\psi_{J,0}} = \frac 1 {\sqrt 2} (
\ket{\ddot\psi_{J,+}} + \ket{\ddot\psi_{J,-}})$, for which
$\norm{\F_0 \ket v} = 1$, lies in $S_{j,+} \oplus S_{j,-} \subseteq
T_\bad$, and the success probability in the bad subspaces could only
be upper-bounded by the trivial $\eta = 1$. This way, all states
from bad subspaces lie inside $T_{t/2,+}$ and $\eta = 1/2$. On the
other hand, we mark the plus subspaces above $j \ge t/2$ as good
instead of bad, because it allows us to prove a stronger bound on
the denominator. We do not lose much in the numerator.

\paragraph{Block-diagonalization of $\Gamma$ and $\O_1$}
Like in the proof of \thmref{thm:tsearch}, we naturally decompose
the basis states $\ket{\ddot\psi_{J,a}}$ with $1 \not\in J$ onto
$\ket{\ddot\psi_{J,a,b}}$ by fixing the first input bit to $b$.  For
the same reasons, some linear combinations of these states lie in
$S_{j,\pm}$ and $S_{j+1,\pm}$.  In particular, if for a $v =
(v_{00}, v_{01}, v_{10}, v_{11})$ we let $\ket{\ddot\psi_J^v}$
denote $v_{00} \ket{\ddot \psi_{J,0,0}} + v_{01} \ket{\ddot
\psi_{J,0,1}} + v_{10} \ket{\ddot \psi_{J,1,0}} + v_{11} \ket{\ddot
\psi_{J,1,1}}$, then $\Gamma \ket{\ddot\psi_J^v} =
\ket{\ddot\psi_J^w}$ with $w = \U G_j \U^* v$ \cite[Claim
17]{asw:symmdpt}, where
\[
\U = \frac 1 {\sqrt 2} \begin{pmatrix}
\alpha_0 & \beta_0 & \alpha_0 & \beta_0 \\
\beta_0 & -\alpha_0 & \beta_0 & -\alpha_0 \\
\alpha_1 & \beta_1 & -\alpha_1 & -\beta_1 \\
\beta_1 & -\alpha_1 & -\beta_1 & \alpha_1 \\
\end{pmatrix} , \qquad
G_j = \begin{pmatrix}
q^j & 0 & 0 & 0 \\
0 & q^{j+1} & 0 & 0 \\
0 & 0 & q^{t/2} & 0 \\
0 & 0 & 0 & q^{t/2} \\
\end{pmatrix} , \qquad
\Z = \begin{pmatrix}
1 & 0 & 0 & 0 \\
0 & -1 & 0 & 0 \\
0 & 0 & 1 & 0 \\
0 & 0 & 0 & -1 \\
\end{pmatrix} ,
\]
and $\alpha_a, \beta_a$ for an $a \in \01$ (and an implicit index
$j$) are defined by \eqnref{eqn:ab} with the threshold
value $t := t-1+a$.  In other words, the columns of $\U$ are
vectors $v$ that put $\ket{\ddot\psi_J^v}$ inside $S_{j,+},
S_{j+1,+}, S_{j,-}, S_{j+1,-}$, respectively.  Furthermore, the
subspaces $S_{j,a,b}$ for $a, b \in \01$, spanned by
$\ket{\ddot\psi_{J,a,b}}$, have the same dimension and there are 3
unitaries that map $\ket{\ddot\psi_{J,0,0}} \to
\ket{\ddot\psi_{J,a,b}}$, hence one can form a complete set of
orthogonal projectors $\Pi = \{ \Pi_{j,\ell} \}_{j,\ell} \cup
\Pi_\triv$ that block-diagonalizes $\Gamma$.  Each projector $\Pi_{j,\ell}$
is 4-dimensional.  $\O_1$ is trivially
block-diagonal in $\Pi$, because $\O_1 \ket{\ddot\psi_{J,a,b}} =
(-1)^b \ket{\ddot\psi_{J,a,b}}$, or, equivalently, $\O_1
\ket{\ddot\psi_J^v} = \ket{\ddot\psi_J^w}$ with $w = \Z v$.

\paragraph{Spectral norm of $\Gamma_1 / \Gamma$ on $\Pi_{j,\ell}$}
Recall $\Gamma_1 = \O_1^* \Gamma \O_1$ and denote $\Gamma^{(j)}
= \Gamma \Pi_{j,\ell} = \U G_j \U^*$ for some $\ell$.  Then
\[
\Gamma_1^{(j)} / \Gamma^{(j)}
= \Gamma_1 \Gamma^{-1} \Pi_{j,\ell}
= \Z \U G_j \U^* \Z \cdot \U G_j^{-1} \U^* .
\]
This matrix is too hard to analyze directly, hence we apply
\corref{cor:multadv2} rather than \corref{cor:multadv}.  Compute
\[
\frac {2 \norm{\Gamma^{(j)} \circ D_1}} {\lmin(\Gamma^{(j)})}
= \frac 2 {q^j} \norm{(\U G_j \U^*) \circ D_1} ,
\qquad \mbox{where }
D_1 = \left( \begin{smallmatrix}
0 & 1 & 0 & 1 \\
1 & 0 & 1 & 0 \\
0 & 1 & 0 & 1 \\
1 & 0 & 1 & 0 \\
\end{smallmatrix} \right) .
\]
Write the matrix $(\U G_j \U^*) \circ D_1$ after swapping the second and third
row and column as $- \frac {q^j} 2 (\begin{smallmatrix}
0 & H_j \\
H_j^* & 0 \\
\end{smallmatrix})$
with
\begin{align*}
H_j &= \begin{pmatrix}
\alpha_0 \beta_0 (q-1) & \alpha_1 \beta_0 (q^{t/2-j} - 1) - \alpha_0 \beta_1 (q^{t/2-j}-q) \\
\alpha_0 \beta_1 (q^{t/2-j} - 1) - \alpha_1 \beta_0 (q^{t/2-j}-q) & \alpha_1 \beta_1 (q-1) \\
\end{pmatrix} \\
&= (q-1) \begin{pmatrix}
\alpha_0 \beta_0 & \alpha_0 \beta_1 \\
\alpha_1 \beta_0 & \alpha_1 \beta_1 \\
\end{pmatrix}
+ (q^{t/2-j}-1) (\alpha_1 \beta_0 - \alpha_0 \beta_1) \begin{pmatrix}
0 & 1 \\ -1 & 0 \\ \end{pmatrix} \\
\norm {H_j}
&\le (q-1) \left\| \begin{pmatrix} \alpha_0 \\ \alpha_1 \end{pmatrix}
    \cdot (\beta_0, \beta_1) \right\|
    + (q^{t/2-j}-1) (\alpha_1 \beta_0 - \alpha_0 \beta_1)
    \left\| \begin{matrix} 0 & 1 \\ -1 & 0 \\ \end{matrix} \right\| \\
\intertext{Use $\alpha_a \le 1$, and $\beta_a \le \sqrt{2 t / n}$ and
$|\alpha_1 \beta_0 - \alpha_0 \beta_1| = O(1/\sqrt{t n})$ for $j \le t/2$
\cite[Claim 19 and 20]{asw:symmdpt}.  Then substitute
$q = 1 + \frac {4 \log(2/\zeta)} t$ and bound $\lambda = q^{t/2} \approx
e^{2 \log(2/\zeta)} = 4 / \zeta^2$.}
\norm {H_j}
&\le 2 (q-1) \sqrt{\frac {2 t} n}
    + (q^{t/2}-1) O \left( \frac 1 {\sqrt{t n}} \right) \\
&= O(1) \cdot \frac {\log (2/\zeta) + 4 / \zeta^2} {\sqrt{t n}}
= O\Big( \frac 1 {\zeta^2 \sqrt{t n}} \Big) .
\end{align*}
We conclude that
\[
\min_j \frac {\lmin(\Gamma^{(j)})} {2 \norm{\Gamma^{(j)} \circ D_1}}
= \Omega(\zeta^2 \sqrt{t n}) .
\]
As we say above, this bound is only valid for $j \le t/2$.  However,
if $j \ge t/2$, then $G_j = q^{t/2} \I$, $\Gamma_1^{(j)} / \Gamma^{(j)} = \I$, and
the analysis in this subspace
is trivial.  This is exactly the reason why we mark the subspaces
$S_{j,+}$ for $j \ge t/2$ as good.

\paragraph{Analysis of the trivial subspaces}
For $j \in \{ t-1, t \}$, the projectors $\Pi_{j,\ell}$ have smaller
dimension than $4 \times 4$, because there are not enough basis
states.  In particular, for $|J|=t-1$, there are only 3 types of
basis states $\ket{\ddot\psi_{J,0,0}} \in S_{t-1,0,0} \subseteq
S_{t-1,0}$, and $\ket{\ddot\psi_{J,1,0}}$ and
$\ket{\ddot\psi_{J,1,1}}$ with $\alpha \ket{\ddot\psi_{J,1,0}} +
\beta \ket{\ddot\psi_{J,1,1}} \in S_{t-1,1}$ and $\beta
\ket{\ddot\psi_{J,1,0}} - \alpha \ket{\ddot\psi_{J,1,1}} \in
S_{t,1}$.  Hence their linear combinations fall into the following
subspaces: $(1,\alpha,\beta) \in S_{t-1,+}$, $(1,-\alpha,-\beta) \in
S_{t-1,-}$, and $(0,\beta,-\alpha) \in S_{t,1} = S_{t,-}$.  Note that
$S_{t,-}$ has a different definition than other $S_{j,-}$, and that there is no subspace
$S_{t,+}$. For, $|J|=t$, the situation is simpler, because there are
only basis states $\ket{\ddot\psi_{J,1,0}} \in S_{t,1,0} \subseteq
S_{t,1} = S_{t,-}$. We conclude that even the projectors onto the
trivial subspaces block-diagonalize $\Gamma$ and $\O_1$.

Now, the actual analysis of the norm of $\Gamma_1/\Gamma$ on these
trivial subspaces is not needed, because $\Gamma = q^{t/2} \I$ on
$\Pi_{t-1,\ell}$ or $\Pi_{t,\ell}$.  We conclude that the norm there
is exactly 1.

\paragraph{Multiplicative adversary bound}
By \corref{cor:multadv2}, using the symmetry over all $i$ and
$\zeta^2 \lambda = \zeta^2 q^{t/2} \approx 4$,
\[
\MADV_{1/2, 4\zeta}(\tthreshold)
\ge \log(\zeta^2 \lambda) \cdot
    \min_j \frac {\lmin(\Gamma^{(j)})} {2 \norm{\Gamma^{(j)} \circ D_1}}
= \Omega( \zeta^2 \sqrt{t n}) .
\]
\end{proof}

\subsection{The OR function}

\newcommand{\OR}{\mathrm{OR}_n}
Let us consider a special case of the $t$-threshold function for $t=1$, the OR function.
It is the decision version of the search function from \secref{sec:search}.  We show
a quadratically better lower bound in terms of the error probability than the one implied
by \thmref{thm:threshold}.

\begin{theorem}
$\MADV_{1/2, 4 \zeta}(\OR) = \Omega(\zeta \sqrt n)$ for $\zeta \ge \sqrt{2 / n}$.
\end{theorem}

\begin{proof}
We use the same subspaces and the same adversary matrix $\Gamma$
like in \eqnref{eqn:gamma-thr}:
\[
\Gamma = \Pi_{S_{0,+}}
    + q \cdot \big( \Pi_{S_{0,-}} + \Pi_{S_{1,-}} \big) ,
\]
$\lambda = q = 2 / \zeta^2$,
and the same block-diagonalization like in the proof of \thmref{thm:threshold}.  However,
we do the analysis more carefully, which is feasible thanks to the fact that we only
have one nontrivial 3-dimensional subspace.  This subspace is spanned by
$\ket{\ddot\psi_{\emptyset,0,0}}$, $\ket{\ddot\psi_{\emptyset,1,0}}$, and
$\ket{\ddot\psi_{\emptyset,1,1}}$.  In this basis, $\Gamma = \U G \U^*$ and
$\Gamma_1 = \Z \Gamma \Z$, where
\[
\U = \frac 1 {\sqrt 2} \begin{pmatrix}
1 & 1& 0 \\
\alpha & -\alpha & \sqrt 2\, \beta \\
\beta & -\beta & -\sqrt 2\, \alpha \\
\end{pmatrix} , \qquad
G = \begin{pmatrix}
1 & 0 & 0 \\
0 & q & 0 \\
0 & 0 & q \\
\end{pmatrix} , \qquad
\Z = \begin{pmatrix}
1 & 0 & 0 \\
0 & 1 & 0 \\
0 & 0 & -1 \\
\end{pmatrix} .
\]
If we applied \corref{cor:multadv2} on this adversary matrix, we
would obtain the same bound as in \thmref{thm:threshold}.  We
instead express $\Gamma_1 \Gamma^{-1} = \Z \U G \U^* \Z \cdot \U
G^{-1} \U^*$ explicitly and get that its eigenvalues are $1$ and
\[
1 + \gamma \pm \sqrt{\gamma^2 + 2 \gamma} ,
\quad \mbox{where }
\gamma = \frac {(q-1)^2} {2 q} (2 \beta^2 - \beta^4) .
\]
We plug in the bound $\beta \le \sqrt{2 / n}$, expand the Taylor series, and obtain
$\norm{\Gamma_1 \Gamma^{-1}} = 1 + \frac {2 (q-1)} {\sqrt{q n}}
+ O(\frac q n)$.  We can neglect the remaining terms when $\zeta \ge \sqrt{2 / n}$.
By \corref{cor:multadv},
\[
\MADV_{1/2, 4 \zeta}(\OR)
    \ge \frac {\log (\zeta^2 \lambda)} {\log (1 + 2 (q-1) / \sqrt{q n})}
    \ge \frac {\log(\zeta^2 q) \sqrt q} {2 (q-1)} \sqrt n
    = \Omega(\zeta \sqrt n) .
\]
\end{proof}

The multiplicative adversary bound for OR is stronger than the additive adversary bound
for polynomially small success probabilities.  Note that success $\frac 1 2 + \zeta$
corresponds to error $\frac 1 2 - \zeta$.  By \corref{cor:addadv}, the additive adversary
bound for OR is
\[
\ADV_{\frac 1 2 - \zeta}(\OR) = \frac {1 - \sqrt{(1-2\zeta) (1+2\zeta)}} 2 \sqrt n
    = \frac {1 - \sqrt{1 - 4\zeta^2}} 2 \sqrt n
    \approx \frac {1 - (1- 2\zeta^2)} 2 \sqrt n
    = \zeta^2 \sqrt n .
\]

\subsection{Designing $\Gamma$ for a general function}

After having presented optimal multiplicative adversary matrices for
several problems, let us make a note on how to design a good $\Gamma$
in general.  It seems that we don't have too much freedom.  All
known good multiplicative adversary matrices $\Gamma$ have the following
structure: $\Gamma$
is a linear combination of projectors $S_j$, where $S_j$ is spanned by
superpositions of input states consistent with fixing exactly $j$ input variables.
This matrix is then simultaneously block-diagonalized with the query
operator.  The diagonal blocks typically overlap with some adjacent subspaces $S_j$
and $S_{j+1}$.  To get a good estimate of the spectral norm of such a block,
the minimal and maximal eigenvalue in this block must not differ too much.
On the other hand, we want the spectral norm of $\Gamma$ be as large
as possible, hence an optimal choice of the multiplicative coefficients seems
to be $q^j \Pi_{S_j}$ for some constant $q$, or more generally
$(\Pi_{i=1}^j q_i) \Pi_{S_j}$, with $q_i$ different in each subspace $S_i$ if the subspaces
for different $i$ have significantly different properties.

The real difficulty seems to lie not in designing good subspaces $S_j$, but in their
combinatorial analysis.

\section{Direct product theorems}
\label{sec:dpt}

In this section we investigate the complexity of evaluating a
function $f$ on $k$ independent instances simultaneously.  We prove
that the multiplicative adversary bound satisfies a \emph{strong
direct product theorem (DPT)}.  Roughly speaking it says that if we
are asked to compute $f$ on $k$ independent inputs in time less than
$k$ times the time for one instance, then the success probability
goes exponentially down.  Ambainis \cite{asw:symmdpt} proved a DPT
for $t$-threshold using these techniques.  Here we show that his
proof actually gives a DPT for any function that has a
multiplicative adversary lower bound.

For a function $f: X \to \Sigma_O$ with $X \subseteq \Sigma_I^n$
and $k \ge 1$, let $f^{(k)}: X^k
\to \Sigma_O^k$ such that $f(x_1, \dots, x_k) = (f(x_1), \dots,
f(x_k))$.  An algorithm succeeds with computing $f^{(k)}$ if all
individual instances are computed right.

\begin{theorem}
For every function $f$ with $\eta \le \frac 1 2$, and $k \ge 361$,
$\MADV_{\eta^{2 k/5}, \zeta^{k/10}}(f^{(k)}) \ge \frac k {10} \cdot
\MADV_{\eta, 4\zeta} (f)$.
\end{theorem}

\begin{proof}
Let $\Gamma, \lambda$ denote the optimal multiplicative adversary matrix
for $f$ with success $\eta$, and its threshold value for good subspaces.  We construct
$\Gamma', \lambda'$ for $f^{(k)}$ as follows \cite[Appendix A.1]{asw:symmdpt}:
\[
\Gamma' = \Gamma^{\otimes k} ,
\qquad \lambda' = \lambda^{k/10} .
\]
We prove that $\max_{i',p} \norm{\Gamma'_{i',p} / \Gamma'} = \max_{i,p}
\norm{\Gamma_{i,p} / \Gamma}$.  This is because, for an $i' = j n + i$,
$\Gamma'_{i',p} = \O_{i',p}^* \Gamma' \O_{i',p}$ with
$\O_{i',p} = \I^{\otimes j} \otimes \O_{i,p} \otimes \I^{\otimes (k-1-j)}$, and thus
$\Gamma'_{i',p} / \Gamma' = \I^{\otimes j} \otimes (\Gamma_{i,p} / \Gamma)
\otimes \I^{\otimes (k-1-j)}$.  Therefore, by \corref{cor:multadv}, if we
choose $\zeta' = \zeta^{k/10}$, the multiplicative adversary bound is
\[
\MADV_{\eta', 4 \zeta'}(f^{(k)}) \ge \frac {\log(\zeta'^2 \lambda')}
    {\log(\max_{i,p} \norm{\Gamma'_{i,p} / \Gamma'})}
= \frac k {10} \cdot \frac {\log(\zeta^2 \lambda)}
    {\log(\max_{i,p} \norm{\Gamma_{i,p} / \Gamma})}
= \frac k {10} \cdot \MADV_{\eta, 4\zeta} (f) .
\]
It remains to analyze the success $\eta'$ of the composed function $f^{(k)}$
in the bad subspaces.

\paragraph{Upper-bounding $\eta'$}
Let $T_\bad, T_\good$ denote the bad and good subspace of $\Gamma$.
For a $v \in \{ \bad, \good \}^k$, let $\ket\varphi \in T_{v_1} \otimes \cdots
\otimes T_{v_k}$ be a product quantum state such that $\norm{\Gamma'
\ket\varphi} < \lambda'$. Since all eigenvalues of $\Gamma$ are at
least 1, if a subspace of $\Gamma'$ corresponds to an eigenvalue
less than $\lambda'$, only less than $k/10$ individual subspaces out
of $k$ can be the good ones. This means that more than
$9 k/10$ instances lie in the bad eigenspace of $\Gamma$ and have thus
success probability at most $\eta$. Since $\ket{\varphi}$ is a product state, the total success
probability of computing all instances right is at most $\eta^{9
k/10}$.  We, however, have to upper-bound the success probability for
all \emph{superposition} states than can come from different combinations
of $T_{v_i}$'s.  In general,
\[
\ket\varphi = \sum_{\substack{v \in \{\mathrm{bad, good}\}^k \\
    |v| < k/10}} \alpha_v \ket{\varphi_v} ,
    \mbox{ where $\ket{\varphi_v} \in T_{v_1} \otimes \cdots \otimes T_{v_k}$
    and $|v|$ denotes \#good subspaces.}
\]
Our assumption about $f$ is that $\norm{\F_z \ket v}^2
\le \eta$ for every $z$ and $\ket v \in T_\bad$.  Thus
\begin{align*}
\norm{(\F_{z_1} \otimes \dots \otimes \F_{z_k}) \ket \varphi}^2
    &= \norm{\sum_v \alpha_v \prod_i \F_{z_i} \ket{\varphi_{v,i}}}^2 \\
&\le \left( \sum_v \alpha_v \prod_i
    \norm {\F_{z_i} \ket{\varphi_{v,i}}} \right)^2 \\
&\le \left( \sum_v |\alpha_v|^2 \right) \cdot \left(
    \sum_v \prod_i \norm {\F_{z_i} \ket{\varphi_{v,i}}}^2 \right) \\
&\le 1 \cdot \eta^{9 k / 10} \sum_{v: |v|<k/10} 1 \\
&= \eta^{9 k / 10} \sum_{i=0}^{k/10} {k \choose i}
\le k {k \choose k/10} \eta^{9 k / 10}
    && {n \choose k} \le \Big( \frac {n e} k \Big)^k \\
&\le k (10 e)^{k/10} \eta^{9k /10}
    && \begin{tabular}{@{}l}
    $\sqrt[5] {10 e} < 2$ \\
    $k \sqrt[5] {10 e}^{k/2} < 2^{k/2}$ for $k \ge 361$
    \end{tabular} \\
&< 2^{k/2} \eta^{k/2} \eta^{2 k/5}
    && \eta \le \frac 1 2 \\
&\le \eta^{2 k /5} .
\end{align*}
Note that in the case of the $t$-threshold function as the base function, the
success probability is $\eta = \frac 1 2$ in both bad (plus) and good (minus)
subspaces, hence we could use a stronger bound $2^{-k}$ instead of $\eta^{9 k /10}$.
There is nothing special about the constant $k \ge 361$; the DPT holds for all $k$,
but we have to take into account the multiplicative factor of $k$ in the success $\eta'$.

We conclude that $\MADV_{\eta^{2 k/5},
\zeta^{k/10}}(f) \ge \frac k {10} \cdot \MADV_{\eta, 4\zeta} (f)$.
\end{proof}

This technique also allows us to prove the direct product theorem when the $k$
instances are distinct functions.

\section*{Acknowledgments}

We thank Andris Ambainis, Peter H\o yer, Sophie Laplante, Troy Lee, and Mehdi Mhalla for fruitful
discussions.

\newcommand{\etalchar}[1]{$^{#1}$}
 \urlstyle{tt} \newcommand{\quot}{\"} \newcommand{\leftbrace}{\{}
  \newcommand{\rightbrace}{\}}

\end{document}